\documentclass[iop]{emulateapj}
\usepackage{color}

\newcommand{\msun}{$\mathrm{M_{\odot}}$}

\def\pasa{\ref@jnl{PASA}} 

\slugcomment{Draft Version \today}

\shorttitle{Progenitors of Planetary Nebulae}
\shortauthors{Badenes, Maoz \& Ciardullo}

\begin{document}

\title{The Progenitors and Lifetimes of Planetary Nebulae}

\author{Carles Badenes\altaffilmark{1}, Dan Maoz\altaffilmark{2}, and Robin Ciardullo\altaffilmark{3}}

\altaffiltext{1}{Department of Physics and Astronomy and Pittsburgh Particle Physics, Astrophysics, and Cosmology Center
  (PITT-PACC), University of Pittsburgh, 3941 O'Hara Street, Pittsburgh, PA 15260, USA; badenes@pitt.edu}

\altaffiltext{2}{School of Physics and Astronomy, Tel-Aviv University, Tel-Aviv 69978, Israel; maoz@astro.tau.ac.il}

\altaffiltext{3}{Department of Astronomy and Astrophysics, and Institute for Gravitation and the Cosmos, The Pennsylvania State
  University, University Park, PA 16802, USA; rbc3@psu.edu}

\begin{abstract}
  Planetary Nebulae (PNe) are amongst the most spectacular objects produced by stellar evolution, but the exact identity of their
  progenitors has never been established for a large and homogeneous sample. We investigate the relationship between PNe and their
  stellar progenitors in the Large Magellanic Cloud (LMC) by means of a statistical comparison between a highly complete
  spectroscopic catalog of PNe and the spatially resolved age distribution of the underlying stellar populations. We find that
  most PN progenitors in the LMC have main-sequence lifetimes in a narrow range between 5 and 8 Gyr, which corresponds to masses
  between 1.2 and 1.0 M$_{\odot}$, and produce PNe that are visible for $27\pm6$~kyr.  We tentatively detect a second population
  of PN progenitors, with main-sequence lifetimes between 35 and 800~Myr, masses between 8.2 and 2.1 M$_{\odot}$, and average PN
  lifetimes of $11^{+6}_{-8}$ kyr. These two distinct and disjoint populations strongly suggest the existence of at least two
  physically distinct formation channels for PNe.  Our determination of PN lifetimes and progenitor masses has implications for
  the understanding of PNe in the context of stellar evolution models, and for the role that rotation, magnetic fields, and
  binarity can play in the shaping of PN morphologies.
\end{abstract}


\section{INTRODUCTION}
\label{sec:Intro}

Despite 250 years of astronomical observations and decades of theoretical work, our understanding of Planetary Nebulae (PNe)
remains rather poor.  In the traditional theoretical picture, the progenitors of PNe are low- to intermediate-mass single stars
\citep[e.g.][]{Abell1966,Balick1987}, which eject a large fraction of their envelope at the end of the asymptotic giant branch
(AGB) phase. Intense photoionization of the ejected material from the exposed stellar core then leads to the formation of an
extended emission nebula. The widely-held belief that the Sun will someday become a PN is rooted in this picture, as is the implication
that all single stars that go through an AGB phase will eventually form PNe. Yet, there is no observational evidence for this. In fact,
the single star hypothesis cannot account for the non-spherical morphologies of many PNe, the observed low rate of PN formation
per unit stellar mass, or the occasional appearance of PNe in old stellar systems like globular clusters
\citep{Jacoby1997,Buzzoni2006,Moe2006,DeMarco2009}.  To explain these and other inconsistencies, new paradigms of PN formation
have emerged, invoking rotation and magnetic fields in single stars \citep{GarciaSegura1999}, the interaction of AGB winds with
close binary companions \citep{Soker1997}, and large mass transfer rates during common envelope episodes
\citep{Ciardullo2005,Nordhaus2007}. All of these mechanisms, or any combination of them, could work in principle, but without
robust connections between PNe and their stellar progenitors, and reliable measurements of mean PN lifetimes, it is difficult to
determine which ones are at play, and what kinds of PNe they produce.  Unfortunately, much of what we know about PNe as a class
comes from either case-by-case studies of individual objects or the analysis of small, heterogenous samples that lack the
statistical rigor needed to probe these issues.

\begin{figure*}
  \centering
  \includegraphics[width=15cm]{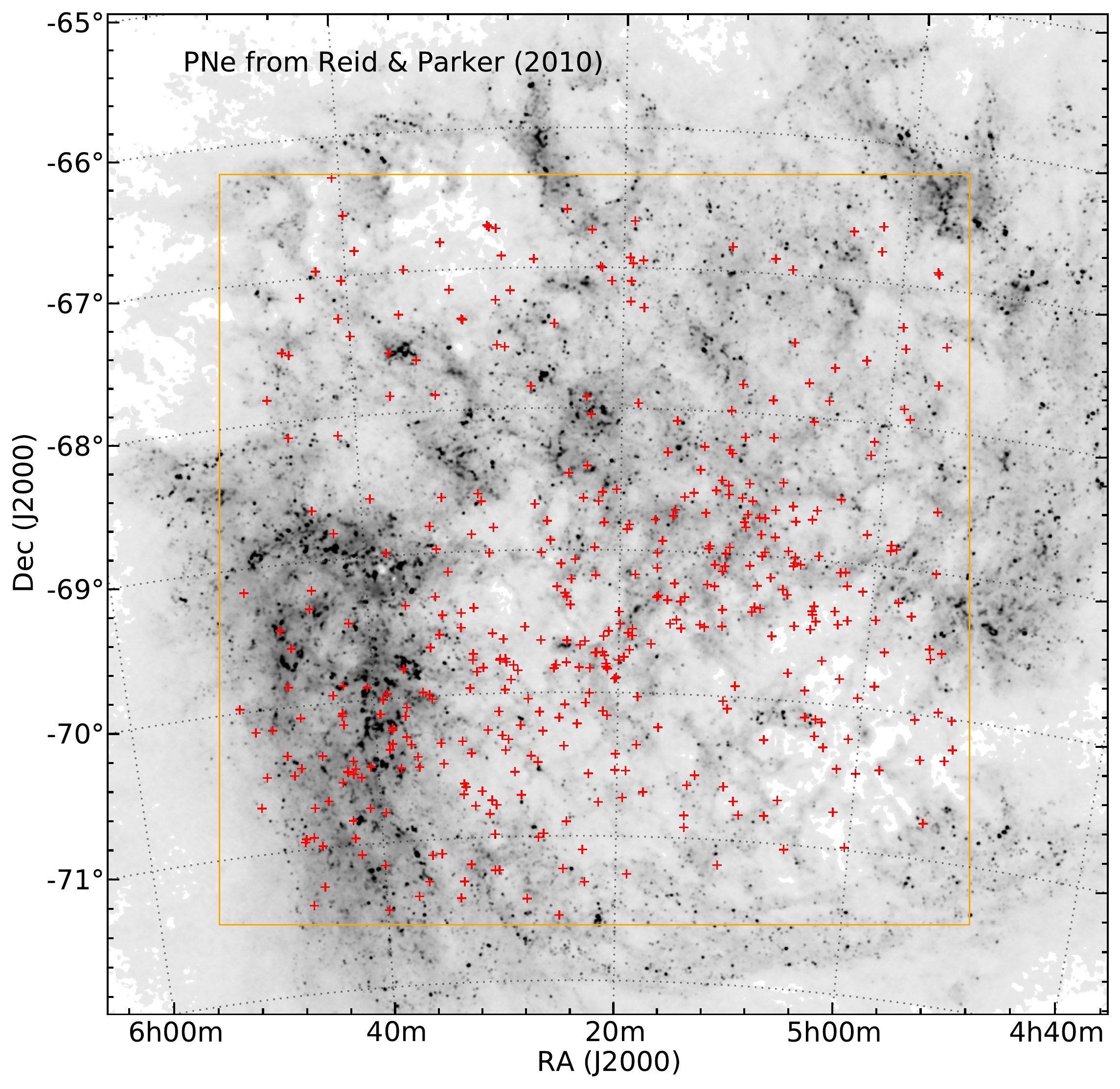} 
  \caption{Planetary Nebulae from \citet{Reid2010} (red crosses), superimposed on an HI map of the
  LMC \citep[][grayscale]{Braun2012}. The area covered by the PN catalog is outlined by the orange box.}
\label{fig:PNLMCMap}
\end{figure*}

Here we present the first statistical analysis of the relationship between PNe and their parent stellar populations in the Large
Magellanic Cloud (LMC). At a well-determined distance of 50~kpc \citep{Pietrzynski2013}, our closest Galactic neighbor is the
ideal setting for this kind of study. On the one hand, the population of PNe in the LMC has been examined in great detail,
culminating in a catalog of 435 spectroscopically confirmed objects that extend more than 7 magnitudes down the [O~III] $\lambda
5007$ PN luminosity function \citep[][see Figure~\ref{fig:PNLMCMap}]{Reid2010,Reid2014}. This catalog of PNe is virtually free of
interlopers, and is statistically complete to a flux limit of $10^{-14}$~erg~cm$^{-2}$~s$^{-1}$ in [O~III] $\lambda 5007$ (i.e., a
line luminosity of $10^{33}$ erg~s$^{-1}$) \footnote{To preserve sample homogeneity and purity, we only use PNe found in field
  MC22 of the UKST survey \citep {Reid2006} whose spectra are classified as `true', and discard `likely' and `possible' objects
  \citep{Reid2014}, as well as outer LMC PNe \citep{Reid2013}.}. On the other hand, the LMC has the best-studied stellar
population in the Local Group. In particular, the stellar age distribution (SAD) has been mapped across the entire galaxy through
a combination of ground-based and \textit{Hubble Space Telescope} observations of millions of individual stars, and their
comparison to theoretical isochrones \citep{Harris2009}. This SAD map consists of 1376 cells covering the inner 64~deg$^2$ of the
galaxy, with a spatial resolution of $12' \times 12'$ ($350 \times 350$~pc at 50~kpc) and a temporal resolution of 0.2 dex (16
bins spanning lookback times between 6 Myr and a Hubble time).

This SAD map can provide unique insights into the properties of the stellar progenitors of PNe. The SAD is the localized version
of the global star-formation history, which gives the amount of stellar mass formed as a function of lookback time for an entire
galaxy. The stars found today in each individual region of the LMC have drifted there over the years, after having been formed at
many different times and locations throughout the galaxy, but the SAD of the region still provides a complete census, within
observational uncertainties, of its current stellar content. As such, it must include the progenitors of all astronomical
transients found in said region (including PNe), provided that the lifetimes of these transients are short compared to the
dynamical drift timescale.

\section{Delay Time Distribution Recovery Method}

From a statistical point of view, the relationship between PNe and their stellar progenitors is encapsulated in the Delay Tine
Distribution (DTD), defined as the rate of production of PNe as a function of time after a hypothetical brief burst of star
formation. Essentially, the DTD is the impulse response, or Green's function, of the formation rate of PNe, and as such it reveals
the evolutionary timescales associated with PN progenitors. Although the DTD can be a key diagnostic for objects with unknown or
uncertain progenitors, such as Type Ia supernovae \citep[see][]{Maoz2014}, it is usually difficult to recover from extragalactic
data sets.  Our work showcases the advantages of deriving DTDs in the Local Group, where we have access to resolved stellar
populations and highly complete object catalogs. For the case at hand, the current ($t=t_0$) PN formation rate, $R_i(t_0)$, in a
certain region $i$ ($i=1 \dots K$) within the LMC is given by the convolution of $\Psi(t)$, the main-sequence turnoff rate of a
single-age stellar population formed at time $t$ (which has units of turnoff stars per year per solar mass formed), with the
region's SAD, $\dot m_i(t)$ (which has units of total mass of stars formed per year):

\begin{equation}
\label{convolution1}
R_i(t_0)=\int_0^{t_0}  \dot m_i(t_0-t) \Psi(t) dt .
\end{equation}

Recall that the SAD includes all stars that are now in region $i$, irrespective of where and when they were formed.  The number of
PNe expected to be observed in the region, $\lambda_i(t_0)$, is the product of the integrand in this expression and the mean
lifetime of PNe, $T_{PN}(t)$, expressed as a function of the main-sequence lifetime of their progenitors $t$:
\begin{equation}
\label{convolution2}
\lambda_i (t_0)= \int_0^{t_0}  \dot m_i(t_0-t) \Psi(t) T_{PN}(t) dt .
\end{equation}
Note that $T_{PN}(t)$ can equal zero if stars of a particular main-sequence lifetime $t$ (i.e., formed with a particular mass) do
not go through a PN phase.  

Given a set of observed PNe in each region $N_{PN,i}$ of the LMC, and an SAD map of the galaxy $\dot m_i(t)$, recovering the DTD
($\Psi(t) T_{PN}(t)$) is a typical inverse problem. To solve it, we follow the procedure described in \citet{Maoz2010}, re-casting
the convolution in Eq.~2 as a discrete sum over time-binned versions of $\dot m_i(t_0-t)$ and $\Psi(t) T_{PN}(t)$:
\begin{equation}
  \lambda_{i} = \sum_{j=1}^{M} m_{i,j} (\Psi T_{PN})_j .
\end{equation}
Here, the number of PNe expected in each spatial cell $i$ of the SAD map, $\lambda_i$ ($i=1 \dots K$) is the product of the
discretized SAD in the cell ($m_{i,j}$, the stellar mass formed during time interval $j$, with $j=1 \dots M$) and the discretized
DTD ($\Psi_{j}$, the number of stars that turn off the main sequence per unit formed stellar mass, per unit time during time
interval $j$, multiplied by the mean lifetime $T_{PN,j}$ of PNe formed from stars with main-sequence lifetimes in the time
interval $j$).  The values of $(\Psi T_{PN})_{j}$ are adjusted with the Markov Chain Monte Carlo solver \texttt{emcee}
\citep{Foreman-Mackey2013}, so that random realizations of a Poisson process with expectation value $\lambda_{i}$ give, on
average, the best fit to the actual number of PNe observed in each cell, $N_{PN,i}$. The value of $N_{PN,i}$ ranges between 0 and
7 in the 732 cells in the SAD map of \citet{Harris2009} which overlap the area of the PN catalog (see Figure~\ref{fig:PNLMCMap}).

Statistical 1 and $2 \, \sigma$ error intervals on each $(\Psi T_{PN})_{j}$ are calculated by applying a highest density criterion
to the posterior probability distributions generated by \texttt{emcee}. To calculate the additional errors due to the uncertainty
in the SADs, we repeat the procedure using the upper and lower limits to the SAD of each cell and temporal bin of the map, and
compute the difference between the new median values and those obtained from the best-fit solution.  Statistical and SAD-related
errors are then added in quadrature. The temporal resolution of the recovered DTD is determined by the size of the observational
sample (i.e., the 435 PNe), as there is a fixed amount of `signal' to be spread amongst the $M$ temporal bins. For the LMC PNe, we
have found that a solution with $M=6$ provides the best compromise between temporal resolution and significance of the detected
progenitor populations, given the statistical and SAD uncertainties.  As a verification test, we also solved for $(\Psi
T_{PN})_{j}$ using the adaptive grid-based search of parameter space with uncertainties based on Monte Carlo simulations described
in \citet{Maoz2010}, and obtained results similar to those described below.

\section{The DTD of PNe in the LMC}
\label{sec:DTD}

\begin{figure*}
\centering
\includegraphics[width=16cm]{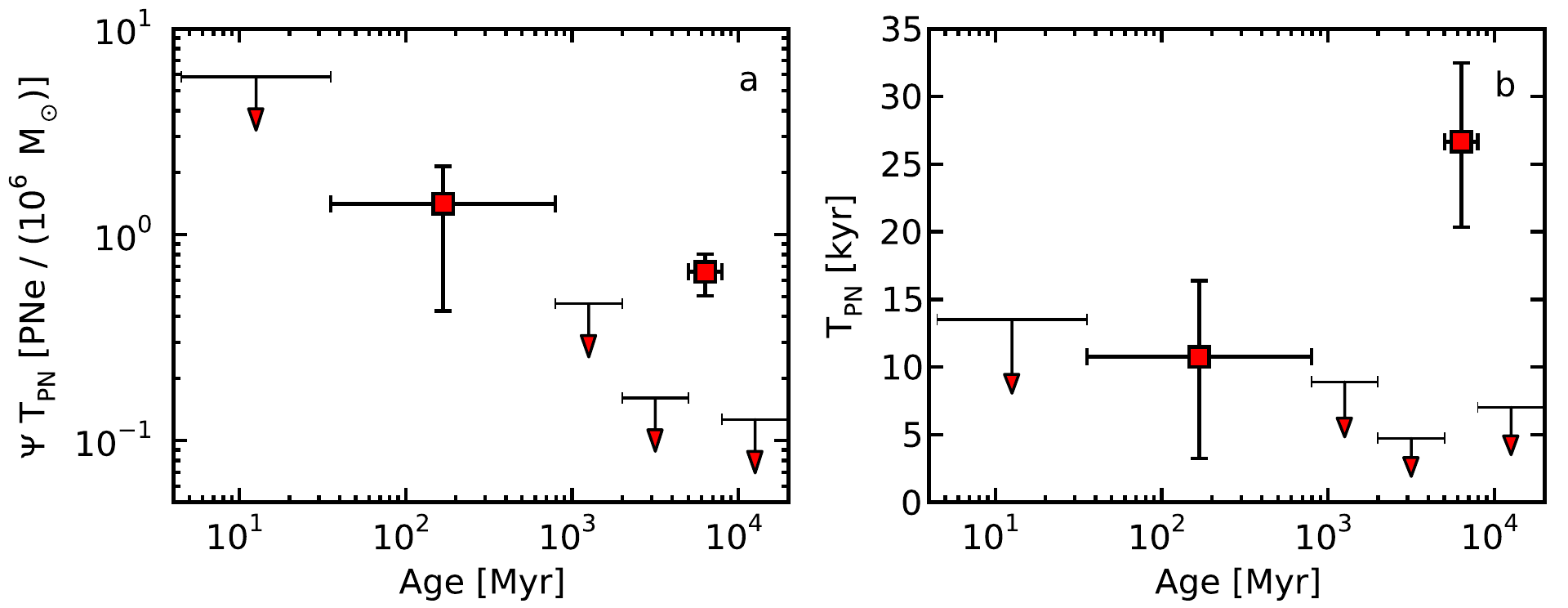}
\caption{(a) Delay time distribution (DTD) of PNe in the LMC, showing the number of PNe produced by a single-age stellar population per unit
  formed stellar mass as a function of time since star formation. There is a clear detection in the $5-8$~Gyr time bin,
  corresponding to main-sequence lifetimes of $1.2-1.0$ M$_\odot$ stars. A second, less significant, detection can be seen in the
  35 Myr to 800~Myr bin, corresponding to stars with initial masses of $2.1-8.2$ M$_\odot$. Upper limits ($2\sigma$) are shown for
  the other bins. (b) Mean PN lifetimes as a function of delay time since star formation, obtained by dividing the DTD in (a) by
  the rate of main sequence turnoff $\Psi(t)$ (Eq.~4). }
\label{fig:LMCPNDTD}
\end{figure*}

\begin{figure*}
\centering
\includegraphics[width=17cm]{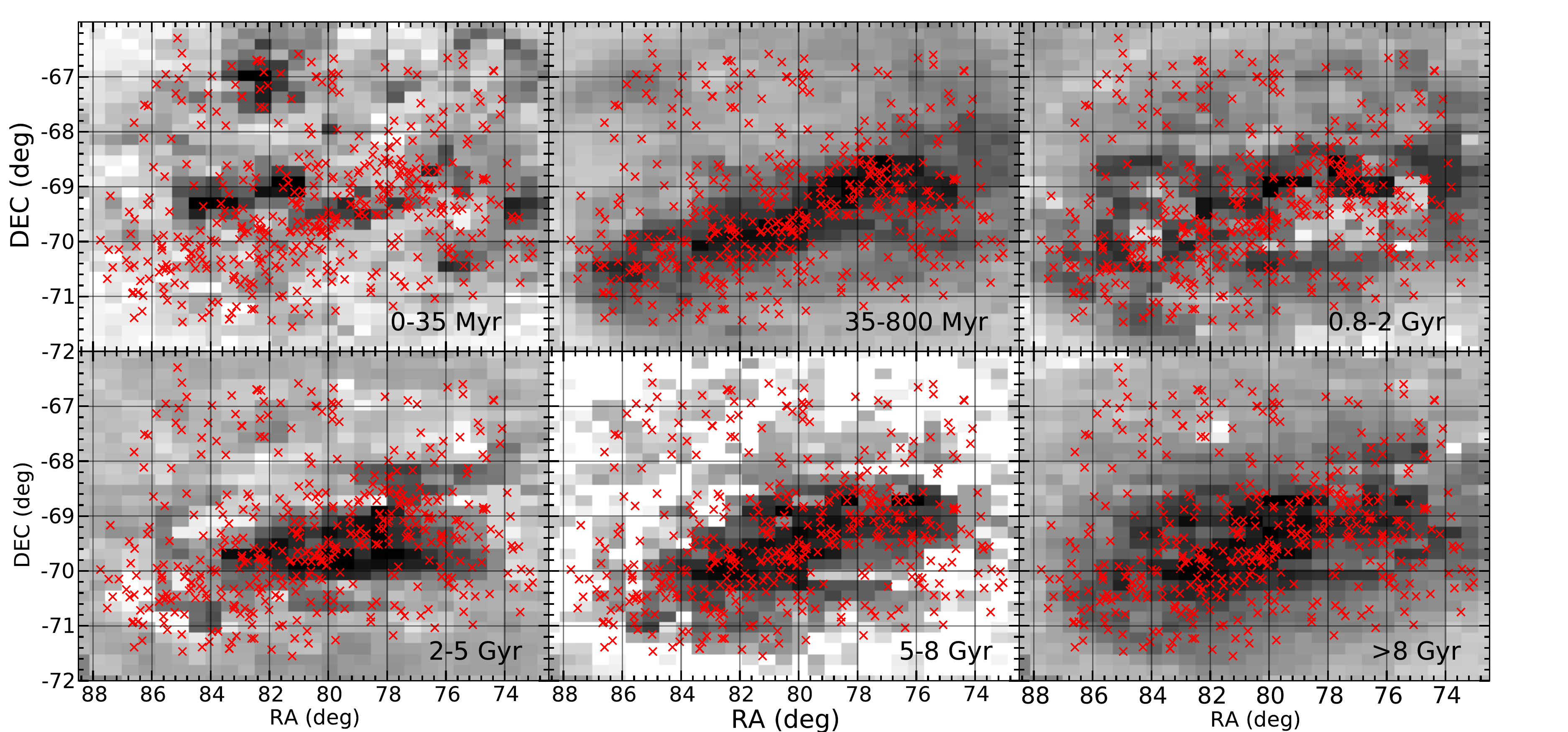}
\caption{PNe from \citet{Reid2010} (red crosses, same in all panels), superimposed on the distribution of stellar mass formed in
  the six temporal bins of the DTD, taken from the SAD maps of \citet{Harris2009} (grayscale). The correlation of the PNe with the
  35-800~Myr and 5-8~Gyr stellar populations, indicated quantitatively by the DTD in Figure~\ref{fig:LMCPNDTD}, is discernible
  qualitatively by eye in the two middle panels.  Most PNe in the LMC come from stars with main-sequence lifetimes in these two
  ranges.}
\label{fig:LMCSADBins}
\end{figure*}

The recovered DTD for the LMC PNe is presented in Figure~\ref{fig:LMCPNDTD} and Table~1.  Its most prominent feature is the clear
detection of a population of PN progenitors with a narrow range of main-sequence lifetimes between $5$ and $8$~Gyr.  These
lifetimes correspond to zero-age main sequence masses between $1.2$ and $1.0$ M$_{\odot}$ in isolated stellar models of LMC
metallicity \citep{Bertelli2008,Bertelli2009}. The DTD also suggests the presence of a second population of progenitors with
main-sequence lifetimes between 35 and 800~Myr (masses between $8.2$ and $2.1$ M$_{\odot}$), but when the SAD uncertainties are
taken into account, the formal significance of this detection drops below $2\,\sigma$. Nevertheless, two separate lines of
evidence lend support to this tentative detection. First, a few well-observed LMC PNe have central core masses above $\sim 0.7$
\msun\ \citep{Villaver2007}, which, through the initial- to final-mass relation of \citet{Kalirai2008}, require progenitors with
main-sequence lifetimes below 800 Myr \citep[in some cases, as low as 60~Myr;][]{Dopita1993}. Second, the spatial correlation
between the distribution of formed stellar mass in each DTD age bin and the locations of PNe in the LMC is strong enough to be
seen by eye in both the 35-800 Myr and the 5-8 Gyr bins (see Figure~\ref{fig:LMCSADBins}).  For all other main-sequence lifetimes,
the correlation is less clear, and we obtain only $2 \, \sigma$ upper limits to $(\Psi T_{PN})_{j}$.

We emphasize that these results do not depend on the details of the global star-formation history of the LMC or the SAD in each
individual region, which is effectively a nuisance function that is disentangled from the DTD in our analysis.  Although dynamical
processes will mix the LMC's stellar populations on timescales of a few disk crossing times \citep[$\sim$200 Myr;][]{Bastian2009},
the measured SAD for each region $i$ gives the actual distribution of stellar ages that have produced the PNe in that specific
location, and therefore the question of when, or where in the galaxy, those stars were formed is irrelevant.

\begin{deluxetable*}{cccc}
  \tablecaption{Delay time distribution of PNe in the LMC \label{tab:DTD}}
  \tablewidth{400pt}
  \tablehead{
    \colhead{Main-sequence lifetime} & 
    \colhead{Stellar Mass \tablenotemark{a}} &
    \colhead{$(\Psi T_{PN})_j$} & 
    \colhead{T$_{PN,j}$} \\
    \colhead{[Myr]} & 
    \colhead{[M$_{\odot}$]} &
    \colhead{[PNe / $10^{6}$M$_{\odot}$]} & 
    \colhead{$[\rm kyr]$} 
  }
  \startdata
  $<35$ & $>8.2$ & $<5.8$ & $< 14$ \\ 
  $35 - 800$ & $8.2 - 2.1$ & $1.4^{+0.7}_{-1.0}$ & $ 11^{+ 6}_{- 8}$ \\ 
  $800 - 2000$ & $2. 1- 1.6$ & $<0.5$ & $< 9$ \\ 
  $2000 - 5000$ & $1.6 - 1.2$ & $<0.2$ & $< 5$ \\ 
  $5000 - 8000$ & $1.2 - 1.0$ & $0.7^{+0.1}_{-0.2}$ & $27\pm6$ \\ 
  $>8000$ & $<1.0$ & $<0.1$ & $< 7$
  \enddata
  \tablenotetext{a}{Stellar masses that correspond to the main-sequence lifetimes listed in the first column for isolated
    stellar models of LMC metallicity \citep{Bertelli2008,Bertelli2009}}
\end{deluxetable*}

The mean lifetimes of the PNe produced in each time bin of the DTD can be calculated by dividing the recovered $(\Psi T_{PN})_{j}$
for each time bin $j$ by the rate at which stars from a coeval population with age $t$ leave the main sequence, $\Psi(t)$. This
rate can be derived as follows. Let us assume a power-law dependence of main sequence lifetime on initial mass, $t_{\rm ms}=t_1
(m/M_{\odot})^\beta$, where $t_1$ is the lifetime, in Gyr, of a 1~M$_{\odot}$ star.  If the mass-normalized initial mass function
is also a power law between mass $m$ and the minimum stellar mass, $m_{\rm min}$ (i.e., $dN/dm= (\alpha+2) (m_{\rm
  min}/M_\odot)^{-(\alpha+2)} (m/M_\odot)^\alpha$), then the main sequence turnoff rate, in stars per Gyr per unit stellar mass
formed, will be
\begin{equation}
  \Psi(t)=\frac{\alpha+2}{\beta}\left(\frac{m_{\rm min}}{M_\odot}\right)^{-(\alpha+2)}
  \left(\frac{t_1}{\rm Gyr}\right)^{-\frac{\alpha+1}{\beta}}\left(\frac{t}{\rm Gyr}\right)^{\frac{\alpha+1}{\beta}-1}
\end{equation} 
From stellar evolution models, $\beta\approx -2.5$ and, for stars with LMC metallicities, $t_1=8$~Gyr
\citep{Bertelli2008,Bertelli2009}. If, for consistency with the SAD maps, we adopt a Salpeter initial mass function
($\alpha=-2.35$, $m_{\rm min}=0.1$ M$_\odot$), we obtain a turnoff rate of $\Psi(t)=0.020 (t/{\rm Gyr})^{-0.46}$ stars per Gyr per
solar mass formed. Dividing $(\Psi T_{PN})_{j}$ by $\Psi_j$ gives the effective lifetime of PNe in each DTD bin.  This lifetime is
a mean over the PN population in a given age bin. If, for example, only half of the stars in an age bin (say, those in close
binaries) actually produce PNe, the lifetimes of those PNe will be twice the value calculated in this way. This method yields
lifetimes of $27\pm6$ kyr for the PNe produced by the older progenitors, and $11^{+6}_{-8}$ kyr for the PNe produced by the
younger ones (see Figure~\ref{fig:LMCPNDTD} and Table~1).  These numbers are in rough agreement with estimates based on a
combination of radiation-hydrodynamics PN models and observed nebular expansion velocities of local objects \citep{Jacob2013}.

\section{Discussion And Conclusions}

The DTD presented here allows us, for the first time, to examine the properties of PN progenitors in a context where the
observational sample of PNe is highly pure and complete, and the entire underlying stellar population has been taken into
account. This has important implications for our understanding of the stellar evolution of low- and intermediate-mass stars. We
confirm the theoretical expectation that some stars do not make a detectable contribution to the LMC PN population, either because
they explode as core collapse supernovae ($t \lesssim 35$~Myr, $M \gtrsim$ 8.0 M$_{\odot}$) or because their post-AGB evolutionary
timescales are longer than the timescale for the dissipation of the ejected envelope \citep[$t \gtrsim 8$~Gyr, $M \lesssim$ 1.0
M$_{\odot}$;][]{Herwig2005}. Surprisingly, we find no contribution to the PN population from stars with main-sequence lifetimes
between 800~Myr and 2~Gyr, which must evolve through an AGB phase of some kind \citep{Herwig2005}. If these stars produce PNe,
they must be either extremely faint (with fluxes below the completeness limits of most PN surveys in the Milky Way), or
short-lived (with mean lifetimes below 9~kyr), or both.

Our DTD and derived PN lifetimes result in an integrated PN formation rate of $\sim 0.02$~yr$^{-1}$ in the surveyed area, which
includes $\sim 80\%$ of the stellar mass of the LMC \citep{Harris2009}. Taking an absolute magnitude of $M_V = -18.4$ for the LMC
\citep{DeVaucouleurs1991} and a bolometric correction of $-0.8$ \citep{Buzzoni2006}, this translates into a bolometric-luminosity
specific PN formation rate of $\sim 7 \times 10^{-12}$~PNe~yr$^{-1}$~L$_{\odot}^{-1}$, which is lower than the typical values
obtained in studies of Galactic PNe. However, Milky Way surveys must deal with heterogeneous, incomplete samples of questionable
purity, and are affected by the systematic uncertainties associated with poorly known distances (\citealt{Parker2006,Sabin2014},
but see \citealt{Frew2015}).  Because all LMC PNe are at a well-determined distance, and have known foreground reddenings, our
analysis is largely free from these problems.

Among the PNe visible in the LMC today, $40^{+23}_{-29}\%$ are generated by the younger progenitors and $25\pm 8\%$ by the older
progenitors.  Thus, roughly one third of the LMC PNe could come from stars in other age bins, for which our DTD only gives upper
limits.  Nevertheless, the presence of two distinct and disjoint progenitor populations strongly suggests the existence of two
separate formation channels for PNe, a possibility that has been qualitatively discussed in the past \citep{Moe2006,
  DeMarco2009,Frew2010}, but never observationally confirmed.  


If we restrict our analysis to the PNe in the brightest third of the sample (L$_{\rm [OIII]}\gtrsim4\times10^{34}$ erg~s$^{-1}$),
the younger progenitor population is not detected.  This suggests, somewhat counter-intuitively, that it is the older,
less-massive progenitors that produce the brighter PNe. If true, this might explain a long-standing issue with the use of the
[O~III] PN luminosity function for calculating extragalactic distances.  This method is well-calibrated in more than a dozen
galaxies with known Cepheid distances \citep{Ciardullo2002}, yet it yields distances to the Virgo and Fornax clusters that are
$\sim 10\%$ smaller than those found with other methods \citep{Jacoby1990,McMillan1993,Freedman2001}.  In the past, this offset
has been attributed to the presence of an unidentified systematic error which only affects galaxies beyond $\sim 10$~Mpc
\citep{Ferrarese2000}, but a more physical explanation is that the brightest PNe in old elliptical galaxies are intrinsically more
luminous than their spiral galaxy counterparts \citep{Ciardullo2012}.  This is consistent with our results.

A detailed interpretation of our DTD in the framework of specific PN formation mechanisms is beyond the scope of the present
work. Nevertheless, we can still put together a basic picture using some additional information. The scale height of bipolar PNe
in the Milky Way is significantly smaller than that of round or elliptical PNe \citep{Phillips2001,Parker2006}, implying that they
have younger progenitors \citep{Corradi1995}. In the LMC, \textit{HST} has imaged 68 of the 435 PNe in our sample, with 49 objects
showing symmetric (round or elliptical) morphologies and 19 showing distinctly bipolar structures \citep{Shaw2006}. The DTD of the
symmetric subsample of PNe shows a significant detection in the 5-8 Gyr bin \citep[as suggested by][]{Stanghellini2009}, while the
subsample of bipolar PNe is too small to produce any statistically significant detections. We have also derived the DTD of the 124
PNe that show enhanced N in their spectra (W. Reid, priv. comm.). These `Type I' PNe \citep{Peimbert1978,Peimbert1983} are
associated with core masses larger than $0.64$ M$_\odot$ \citep{Kaler1990}, corresponding to main-sequence masses above $\sim$2.25
M$_\odot$. The DTD shows a marginal (1.7$\sigma$) detection in the 35-800 Myr bin, consistent with this minimum core mass. Taken
together, these results suggest a scenario wherein the 35-800 Myr progenitors produce fainter, more asymmetric PNe, possibly of
binary origin \citep{Nordhaus2007}, while the 5-8 Gyr progenitors produce brighter, more symmetric PNe, perhaps via traditional
single star evolution.

This basic picture is in rough agreement with the properties of individual PNe in the LMC \citep{Dopita1993,Villaver2007}, and the
Galactic Bulge \citep{Gesicki2014}, but a rigorous statistical validation would require a complete morphological census. The main
features of our proposed scenario are also supported by theoretical work. On the one hand, the PN lifetimes stemming from the
older progenitors are consistent with state-of-the-art radiation-hydrodynamics models for single-star PNe \citep{Jacob2013}. On
the other hand, it has been shown that binary systems that enter a common envelope phase while the primary is on the AGB can form
bipolar PNe, provided that the primary has a minimum mass of $\sim$ 2 M$_\odot$ \citep{Soker1998}. We leave a more detailed
evaluation of these and other specific PN formation mechanisms to future work.

We conclude by pointing out that the methods and techniques presented here are applicable to a wide variety of astronomical
objects, from variable stars (Cepheids, RR Lyrae, $\delta$ Scuti, etc.) to interacting binaries (novae, cataclysmic variables,
high- and low-mass X-ray binaries, etc.).  The best available catalogs for these objects, and the most complete census of their
progenitor stellar populations, are found in Local Group galaxies. The recovery of high-quality DTDs from these data sets can
provide much-needed tests for specific stellar evolution scenarios.

\acknowledgements We acknowledge useful discussions with Avishay Gal-Yam, Dan Foreman-Mackey, Dustin Lang, and Dennis
Zaritsky. Letizia Stanghellini and Warren Reid kindly provided us with updated lists of PN morphologies from \textit{HST} and PNe
with N-rich spectra for the revised version of the paper, which greatly benefitted from the suggestions of the anonymous referee.
DM acknowledges support by the I-CORE Center of the Planning and Budgeting Committee and the Israel Science Foundation.  The
Institute for Gravitation and the Cosmos is supported by the Eberly College of Science and the Office of the Senior Vice President
for Research at the Pennsylvania State University.

\end{document}